\pdfoutput=1 
\RequirePackage{ifpdf}
\documentclass{JINST}

\usepackage[subrefformat=parens,labelformat=parens]{subfig}
\usepackage{lineno}

\title{Development and Characterization of a Multi-APD  Xenon Electroluminescence TPC}

\author{T. Lux$^a$\thanks{Corresponding author.}~,
				A. Garcia Soto$^a$,
				O. Ballester$^a$,
				S. Bordoni$^a$,		
				I. Gil-Botella$^b$,
				N. Hamer$^b$,
				J. Illa$^a$,		
				G. Jover~Ma\~{n}as$^a$$^d$,				
				C. Mart\'{i}n-Mar\'{i}$^a$,
				C. Palomares$^b$, 
				J. Rico$^a$,
				F. Sanchez$^a$,
        R. Santorelli$^b$ and 
				A. Verdugo$^b$.\\
\llap{$^a$}Institut de F\'{i}sica d\'{}Altes Energies (IFAE),\\ Edifici Cn, Universitat Aut\`{o}noma de Barcelona,  Bellaterra (Barcelona), Spain\\
\llap{$^b$}Centro de Investigaciones Energ\'{e}ticas, Medioambientales y Tecnol\'{o}gias (CIEMAT),\\ Madrid, Spain\\
\llap{$^d$}ALBA-Cells,\\ Barcelona, Spain\\

E-mail: \email{Thorsten.Lux@ifae.es}}

\abstract{The performance of an electroluminescence (EL) Time Projection Chamber (TPC) with a multi avalanche photodiode (APD) readout was studied in pure xenon
at 3.8 bar. Intercalibration and reconstruction methods were developed and applied to the data yielding energy resolutions as good as 5.3$\pm$0.1 \% FWHM for 59.5 keV gammas from $^{241}$Am. This
result was reproduced with a Monte Carlo (MC) based on Geant4 and Penelope which predicted 5.2 \% FWHM for the used setup.
Point resolutions of $\approx 0.5$ mm were obtained with a pitch of 15 mm between the APDs. These results show that multi-APD readout is a competitive technology for EL detectors filled with pure xenon.}


\begin{document}

\section{Introduction}
Over the last decade electroluminescence (EL) xenon time projection chambers (TPCs) became an appealing detector concept for various particle physics experiments. 
The advantages of a an EL TPC include  high gain, excellent energy resolution and low detection threshold.  In a pure noble gas at high pressure. - e.g. at 15 bar \cite{oliveira2013results} - three-dimensional reconstruction of the physics process can be achieved. 
A readout technology  with these features is interesting for dark matter \cite{Aprile:2005ww} and double beta decay experiments \cite{Granena:2009it}. 
In addition to their possible applications to  particle physics xenon EL detectors may provide powerful gamma cameras for astronomy, homeland security and medical applications. In the latter field they have potential as Single-Photon-Computerized-Tomography (SPECT) detectors. \\   
In an EL detector light is produced by two physical processes. Primary scintillation light is directly generated when a charged particle traverses the detector or a X-ray converts within the gas, ionizing and exciting the gas molecules. About 9 photons per keV of deposited energy \cite{1748-0221-3-07-P07004} are emitted in the de-excitation process. The primary ionization electrons drift in a relatively low electric field to a region of high electric field of a few kV/cm - defined by two wire meshes Electrons entering this region are accelerated to energies that allow to excite the gas molecules without ionizing it. In the de-excitation process photons are emitted isotropically. This light is referred to as secondary scintillation or EL light. Xenon EL light has a wavelength of 172 nm, in the vacuum ultra violet (VUV), just as the primary scintillation light. 
Excellent energy resolutions close to the photo statistics limit can be achieved because the EL amplification process is linear. These resolutions are comparable to those achieved with solid-state CdTe detectors.\\
Photomultiplier tubes (PMTs) are currently chosen to detect the photons emitted by xenon. They  have the well-known advantages of high gain and low noise,  accompanied by significant drawbacks: their bulk, which limits their trackink resolution, and their  their fragility in high-pressure environments.
Solid-state detectors such as Avalanche Photo Diodes (APDs) and Multi Pixel Photon Counters (MPPCs, also known as SiPMs) may overcome these disadvantages, for they can be produced in sizes small enough for precise tracking and they are resistant to high pressures. APDs are solid-state detectors based on silicon pn-junctions. When a electron-hole pair is created in the depletion region and the voltage applied to the pn-junction is high enough, of the order of 10$^5$ V/cm, an avalanche process starts in which more electron-hole pairs are produced. A MPPC consists of a pixelized  array of APDs operated in Geiger mode, with pitches of up-to 100 $\mu$m. While in the Geiger mode gains of 10 $^5$ can be obtained,, sufficient to detect single photons, in this mode of operation one cannot avoid non-linearities at light intensities at which there is a high probability that in a than one photon is absorbed in one APD pixel. The available sizes of SiPMs used to be 1x1 mm$^2$ and now are 3x3 mm$^2$. This would require a large amount readout channels to achieve a large coverage in the readout plane.

In this work we read out the EL signal using APDs, because our goal was to precisely measure the energy deposited in the gas The APDs were  operated at voltages at which the amplification is proportional to the number of photons impinging on the sensitive area. The main disadvantage in this regime is lack of sensitivity to single photons because the gain is limited to a few hundred times, not sufficient to overcome the intrinsic noise. The APDs are commercially available in sizes of 5x5 and 10x10 mm$^2$ which allow interesting imaging applications. 
Nevertheless we also consider MPPCs coupled to a wavelength shifter as described in reference \cite{1742-6596-309-1-012004} as an interesting solution. A combination of the two readout technologies could be an excellent choice for high-pressure Compton Cameras. However, we leave the idea of using MPPCs for the primary scintillation detection for the future and focus here on energy and track measurements by EL readout using  APDs.\\
Several groups have been studying the performance of EL detectors using single sensors \cite{Moszynski2000230,Moszynski2003226,Monteiro:2007vz,MonteiroP06012,Monteiro2005,P09010,P08005} and demonstrated that similar energy resolutions can be achieved with PMTs and APDs \cite{940072}. The next logical development is to show that good energy resolutions are also achievable with multi-sensor readouts. The first results presented by our group with a small setup of five densely packed APDs (pitch: 10.3 mm) showed excellent performance, with 7.7 \% FWHM for 22.1 keV photons and a detection threshold of about 3 keV at 1.65 bar \cite{Lux:2011vp}. \\
In this paper we present the results obtained with a larger setup of 25 APDs operated in xenon at 3.8 bar. The aim of the study presented here was to prove that multi-APD readouts could provide an excellent energy and spatial resolution  theby making this readout technology suitable for a large range of applications. For this purpose appropriate simulation and analysis tools also had to be developed. \\
The paper is structured as follows: In sec.~\ref{sec_setup} the experimental setup is described. Sec.~\ref{sec_analysis} contains the analysis strategy. The results of the measurements are presented in sec.~\ref{sec_results} and in sec.~\ref{sec_mc} the expected performance and extrapolations based on the measurements are discussed. Finally the conclusions are presented in sec.~\ref{sec_conclusions}.

\section{Experimental Setup} \label{sec_setup}
\begin{figure}[!t]
\centering
\centering
\includegraphics[width=4in]{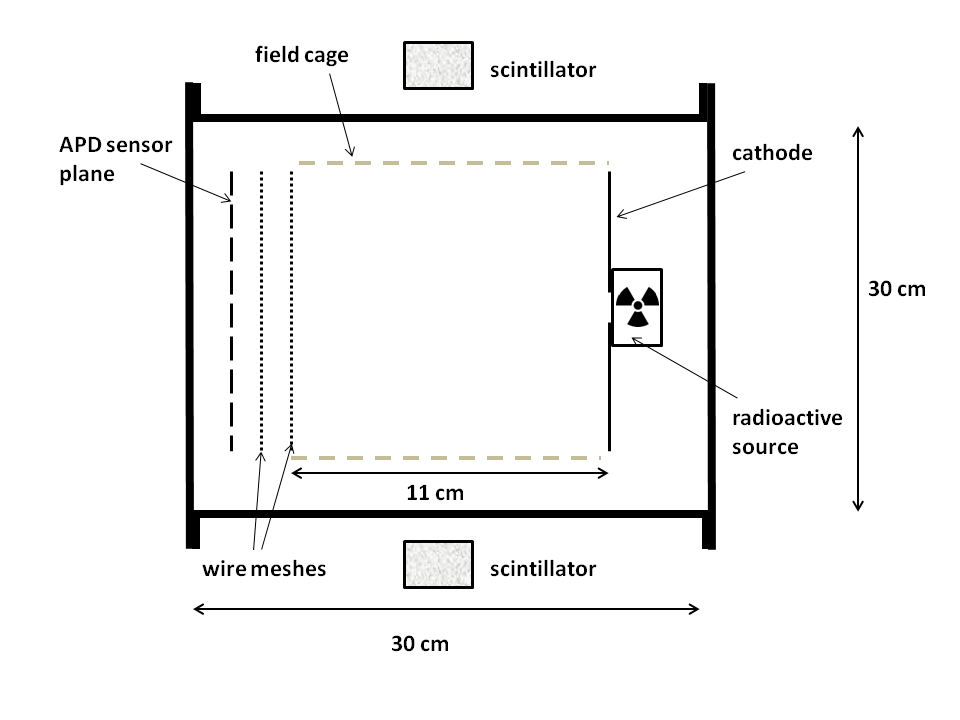}
\caption{Unscaled sketch of the detector. }
\label{fig_sketch}
\end{figure}

\begin{figure}[!t]
\centering
\centering
\includegraphics[width=4in]{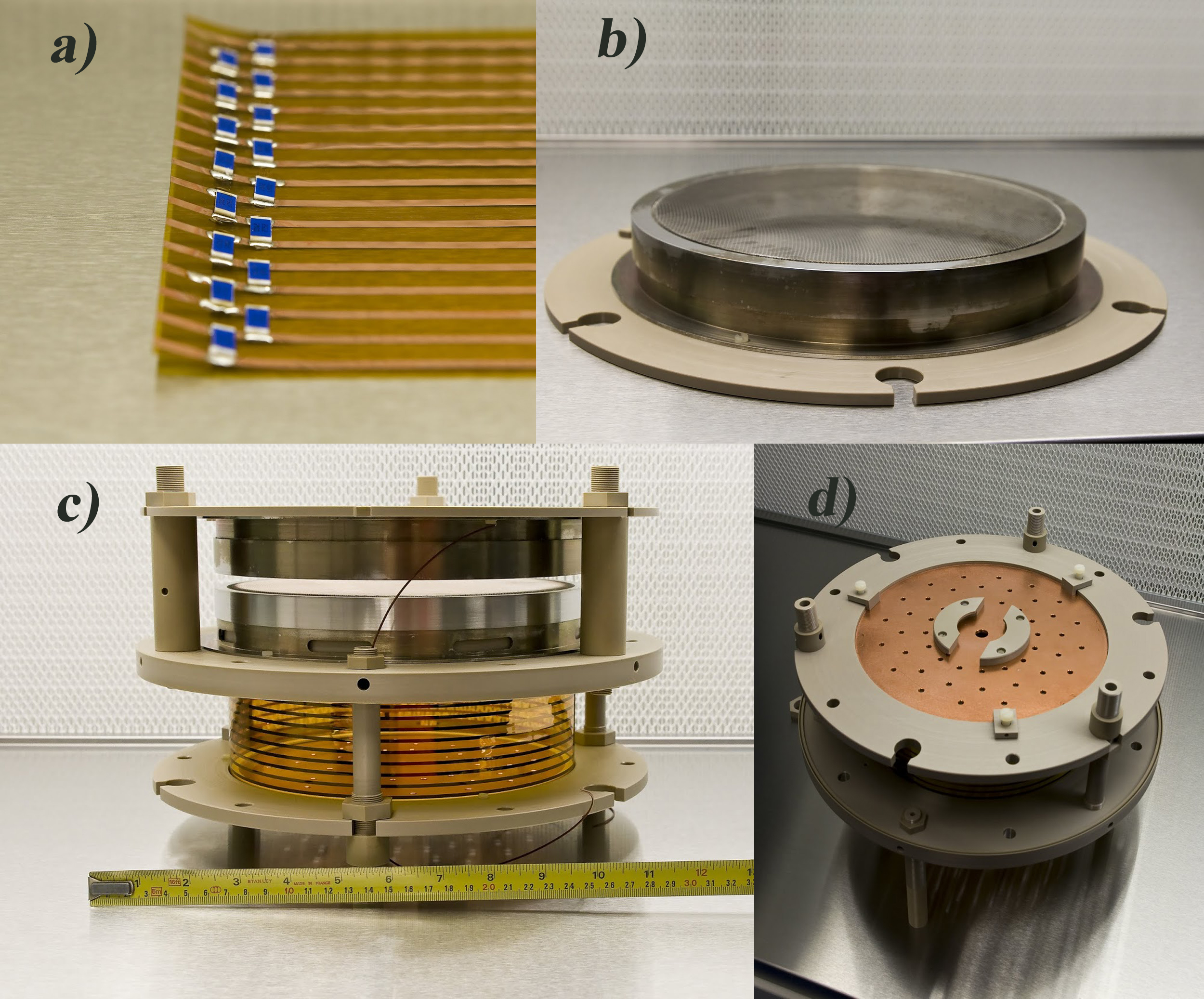}
\caption{ a) shows the copper-coated Kapton foils,  b) shows one of the EL meshes, in c) the complete field forming structure is shown and in d) is shown the cathode with the radioactive source holder. }
\label{fig_inner}
\end{figure}

\begin{figure}[!t]
\centering
\includegraphics[width=3.5in]{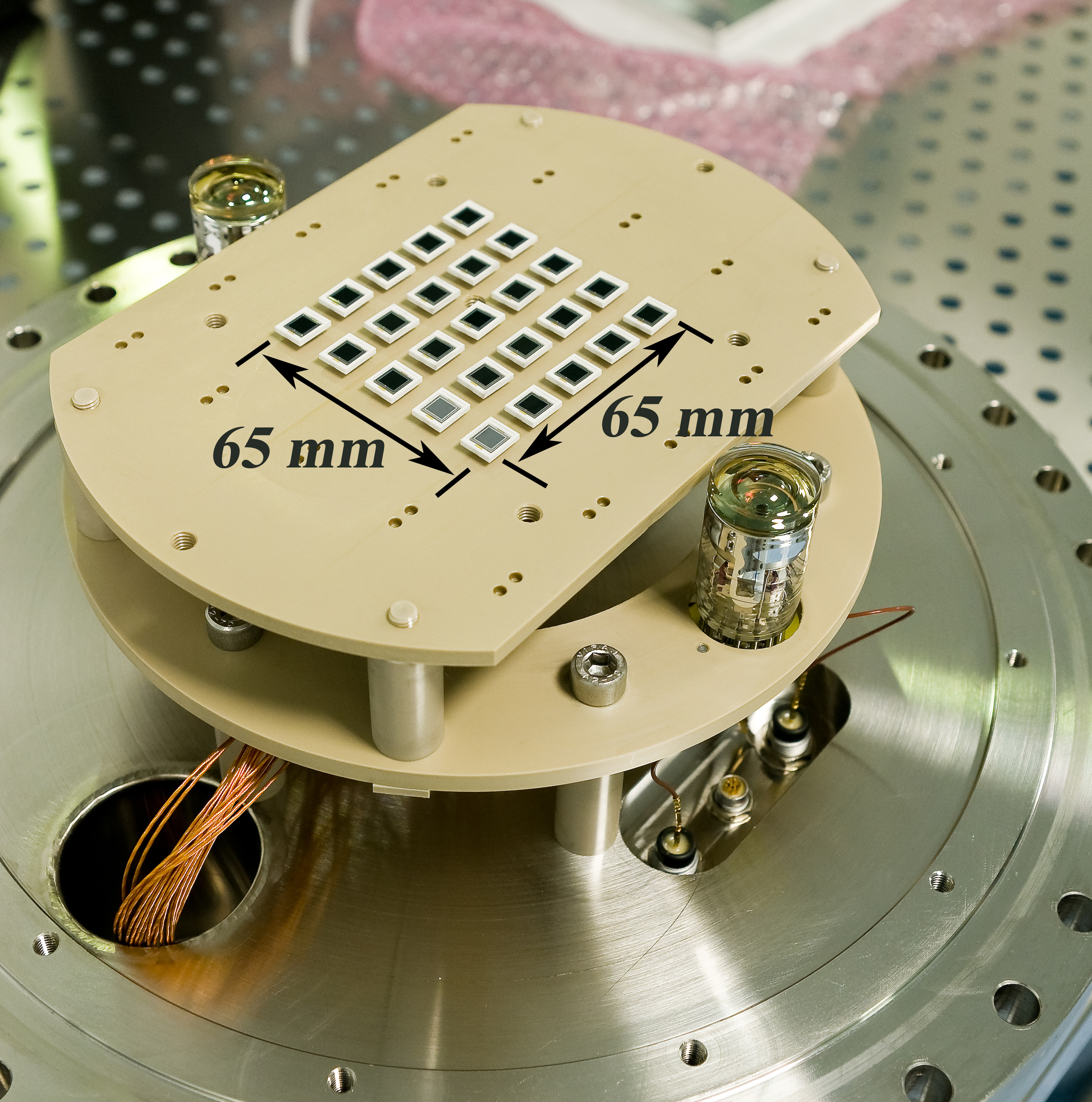}
\caption{The readout plane with 25 APDs and the two PMTs. }
\label{fig_readoutplane}
\end{figure}

The apparatus built for this study consists of an EL TPC designed to operate  up to 5 bar together with the appropriate gas and vacuum systems. 
The stainless steel pressure vessel is cylindrical, with height and diameter of about 30 cm (Fig.~\ref{fig_sketch}). For tracking, two plastic scintillators of 2 and 3 cm width are installed above and below the vessel allowing to trigger on cosmics crossing the chamber. Inside the vessel the field cage structure consisting of the cathode, the field-forming structure and the two EL meshes is installed. The field cage has a diameter of about 20 cm and a drift path between cathode and the first EL mesh of about 11 cm. For this study the gap between the two EL meshes was chosen to be 7 mm. The EL meshes are made of stainless steel wires with diameter of 80 $\mu$m and a pitch of 900 $\mu$m providing $\approx$80\% optical transparency. Special care was taken in the design of the EL meshes to ensure that the maximal electric field strength is achieved between the meshes and not between the frames tensing them. This precaution is useful because the operation of this chamber at high pressures requires high voltages of up-to 20 kV at the cathode and 15 kV at the first EL mesh. We also custom-designed HV feedthroughs to stand these voltages. The field-forming structure guarantees a good electric field homogeneity in the drift region: it consists of a copper-coated Kapton foils between which 5 MOhm resistors are soldered. In the center of the cathode, made of oxygen-free copper, radioactive sources can be installed. The inner structure is shown in Fig.~\ref{fig_inner}.
For the study presented here we used a $^{241}$Am source ($\approx$5~kBq). This source provides several of gammas lines, the most significant ones being at 13.95, 17.75, 20.78, 26.3 and 59.5 keV \cite{Lepy2008715,Gunnink1976}. In xenon several additional escapes peaks in the range between 26 and 34 keV can be seen, due to {\it{K}} shell X-ray fluorescence lines (29.46, 29.78, 33.59 and 34.42 keV \cite{507079}), with the most pronounced one around 30 keV. The results presented here are from the line at 59.5 keV, unless explicitly stated otherwise. The $\alpha$s also emitted from the source were shielded out. \\
For all support structures of the field cage and the HV feedthroughs PEEK$^{\circledR}$ was used due to its good insulation and low outgassing properties. 
Due to the high price of xenon, special care was taken to minimize gas losses and to ensure good gas purity during data taking. 
For this reason the pressure vessel was evacuated to a residual pressure of about 10$^{-5}$ mbar before filling it with xenon thereby minimizing contamination with air. To ensure gas purity during the operation of the detector, a getter (ST172 from SAES) was installed in the system. The gas losses were minimized with a recovery system based on condensation of xenon in a gas bottle attached to the gas system  immersed in a LN$_2$ bath. This process is also periodically used to clean the gas by exploiting the fact that most impurities condense at significant lower temperatures than xenon. The procedure was repeated before each data taking to maximize gas quality.\\
The readout plane is placed about 4 mm behind the second EL mesh. It contains 25 APDs of which 24 are read out (Fig.~\ref{fig_readoutplane}). The APDs have an active area of about 5x5 mm$^2$ and are arranged in the $xy$ plane as a 5x5 matrix with a pitch of 15 mm between the APDs ( $\approx$ 11\% coverage). The APDs are standard devices from Hamamatsu (S8664-SPL) which have been made sensitive to VUV photons by removing the protective layer covering the sensitive area.  Their quantum efficiency for 172 nm photons is high, being about 70\%. These devices were characterized in a previous study \cite{Lux201211}. Each APD is independently powered by a ISEG VDS181 24-channel HV power supply and its nominal operation voltage is chosen based on the data sheet of the producer and the noise performance of the APD. \\
The signal of each APD is first amplified and shaped by custom-made electronics before being digitized by a 65 Mbit/s ADC V1740 from CAEN.
This  electronics is based on a transimpedance amplifier which converts the output current at the anode of the APD to voltage and amplifies it one step. The highest gain of the amplifier is about 10$^5$, obtained without reducing the amplifier bandwidth below the signal bandwidth, which allows collecting the whole signal.  Following the first amplifier stage a semi-Gaussian shaper filters the signal eliminating high and low frequency noise. The signal after the shaper is a few $\mu$s wide, corresponding to the typical EL signal time spread. Each preamp/shaper board contains 8 electronic channels.
The DAQ system to read the digitized data from the ADC is based on the DAQ framework MIDAS \cite{MIDAS}. Three different event trigger modes were implemented. The first mode is a random software trigger which is used to determine the noise and pedestal of each APD during data taking. The second trigger mode introduces a threshold on the ADC signal and is used for the data taken with the $^{241}$Am source. Due to the characteristics of the ADC the trigger condition is fulfilled when in one of a group of 8 APDs the signal exceeds the threshold. The threshold is chosen either at a low value, setting it at the level of the most noisy APD (in order to measure the full signal spectrum), or at high value, such that only the high energy gamma lines of about 30 and 59.5 keV are detected. This trigger mode provides the position in the $xy$ plane but not the prompt time $T0$ or $z$ - reference of the signal, for the primary scintillation light from $^{241}$Am gamma source signal, about 540 photons emitted in 4$\pi$, is too weak to be detected with the APDs. Finally, for data-taking with cosmics an external trigger based on the coincidence signal from the two scintillators is used. This allows operating the detector in 3D mode as a time projection chamber (TPC).



\section{Data Analysis}
\label{sec_analysis}

The data analysis consists of three main steps: first, extracting the signal from the APDs; second, applying an intercalibration to equalize the APD responses; third, determining the deposited energy and the interaction position within the chamber using the signal distribution in the APD plane. The standard operating conditions are summarized in Table~\ref{conditions}.

\begin{table}[h!!]
\begin{center}
\begin{footnotesize}
\begin{tabular}{| c | c | c | c | c |}
\hline
Pressure & $3.8~bar$ \\
\hline
$E_{EL}$ & $4~kVcm^{-1}bar^{-1}$ \\
\hline
$E_{drift}$ & $100~Vcm^{-1}bar^{-1}$ \\
\hline
$\overline{V}_{APD} $& $398~V$ \\
\hline
\end{tabular}
\end{footnotesize}
\end{center}
\caption{Standard conditions during data taking. $E_{EL}$ is the reduced electric field applied between the two meshes, $E_{drift}$ the reduced electric field 
in the drift region and $\overline{V}_{APD}$ the average APD bias voltage. }
\label{conditions}
\end{table}

\subsection{Signal treatment}
\label{SignalTreatment}

\begin{figure}[h!!]
\centering
\includegraphics[width=0.5\textwidth]{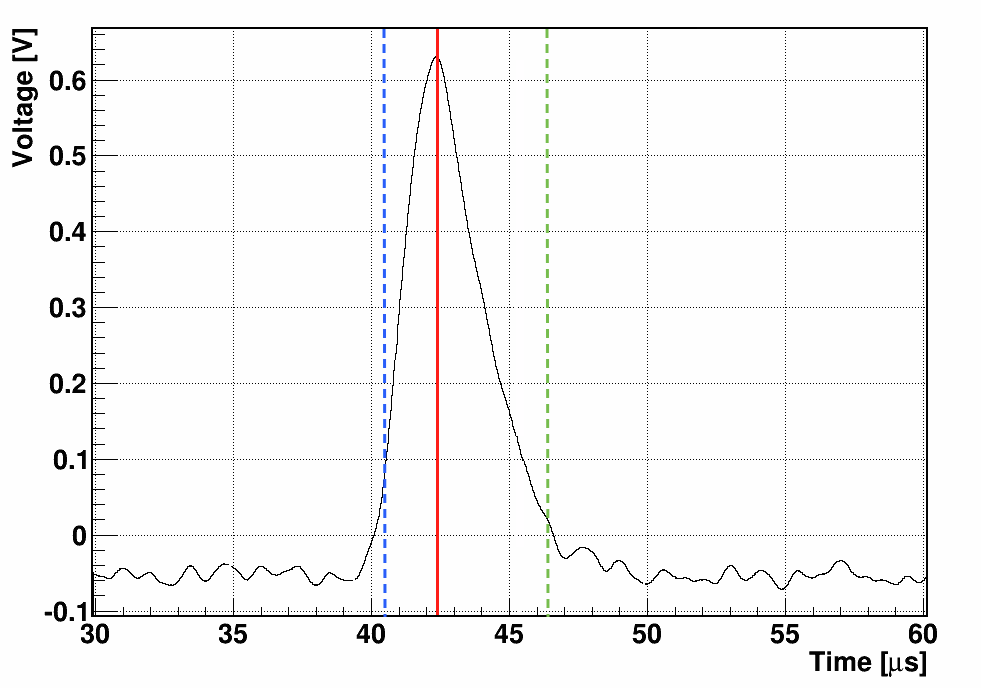}
\caption{Typical waveform from the APD with highest amplitude. The solid red line defines the time of the maximum is reached. The dashed lines indicate the times $t_{up}$ (green) and $t_{low}$ (blue) chosen for the integration window to optimize the energy resolution. }
\label{waveform}
\end{figure}

In the first step the waveform of each APD was converted into a value representing the energy deposited in this APD, which is proportional to the overall energy of photons impinging on it. For this step a time window was defined around the peak time position of the APD with the highest amplitude. The waveform was integrated for each APD separately and the obtained value was divided by the number of time bins. Fig.~\ref{waveform} shows the typical waveform from the APD with the highest amplitude.

The integrated signal of each APD, $E_i$, is calculated as follows. The width of the time window (defined by $t_{low}$ and $t_{up}$) was optimized to achieve the best energy resolution. A stable value of the energy resolution was found between $1.0~\mu s<t_{low}<2.0~ \mu s$ and $1.5~\mu s<t_{up}<4.1~\mu s$ with variations of less than $0.1\%$ within this range. From this stable range, we picked the time window that contains the whole signal, $t_{low}=1.8~\mu s$ and $t_{up}=4.1~\mu s$, which yields an energy resolution of $9.5\%$.

\begin{figure}[h!!]
\centering
\includegraphics[width=0.5\textwidth]{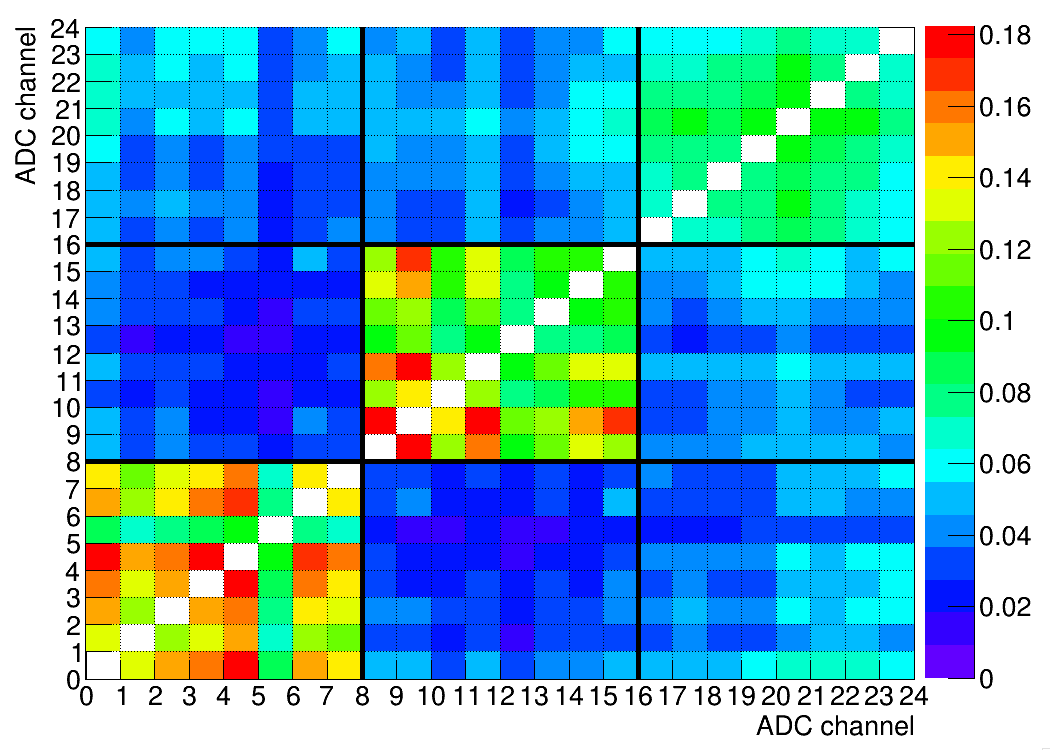}
\caption{Correlation matrix between ADC channels. Solid black lines separate the different preamp/shaper boards.}
\label{correlation}
\end{figure}

The APD pedestals were determined as follows. Every 1000 events triggered by the threshold condition, 100 events were taken with the random software trigger. The waveforms from these events were treated exactly as the real data. The obtained distributions have a Gaussian shape, providing the pedestal (mean value) and noise (sigma) for each APD. The mean values were subtracted from the waveforms. The correlations between the pedestals of the different channels were taken into account. In Fig.~\ref{correlation} one can appreciate strong correlations within groups of 8 ADC channels, corresponding to the channels belonging to the same preamp/shaper board (see sec.~\ref{sec_setup}).

\subsection{APD intercalibration}
\label{APDcalibration}

Intercalibration of the APD gains is critical because they depend exponentially on the applied bias voltage and not all APDs are operated exactly at the same gain. For this purpose we developed a method based on the responses to the $59.5$~keV peak of the $^{241}$. It is based on two fundamental assumptions: 1) the distribution of the EL light detected in the readout plane has radial symmetry and 2) the amount of EL photons produced is position independent. 

\begin{figure}[h!!]
\centering
\includegraphics[width=0.5\textwidth]{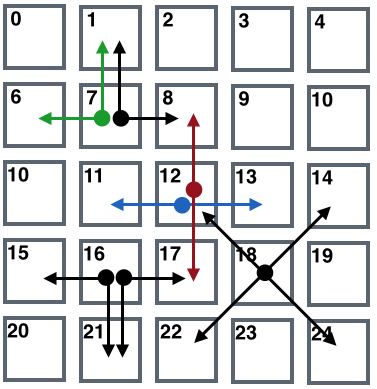}
\caption{Schematic view of the APD plane showing examples of the 8 possible asymmetry couplings $A_i^{j;k}$. The number within the boxes is the index of each APD, the arrows point to APD$_j$ and APD$_k$ and the big dots define APD$_i$. For example: red=$A_{12}^{7;17}$; blue=$A_{12}^{11;13}$; green=$A_6^{5;1}$.}
\label{apd_hor}
\end{figure}

Two factors (referred to as $T$ and $H$) are relevant for this method. To compute them it is necessary to define a variable called {\it Asymmetry}, $A_i^{j;k}=\frac{E_{j}-E_{k}}{E_{j}+E_{k}}$, which computes the light sharing between APD$_j$ and APD$_k$ when the maximum integrated signal is at APD$_i$ with the constraint that the APDs$_{j,k}$ are at identical distances to APD$_i$. Fig.\ref{apd_hor} shows the 8 possible pairs of APD$_{j,k}$ defined under this condition.

\begin{figure}[h!!]
\centering
\subfloat[]{\label{asy_cut} \includegraphics[width=0.48\textwidth]{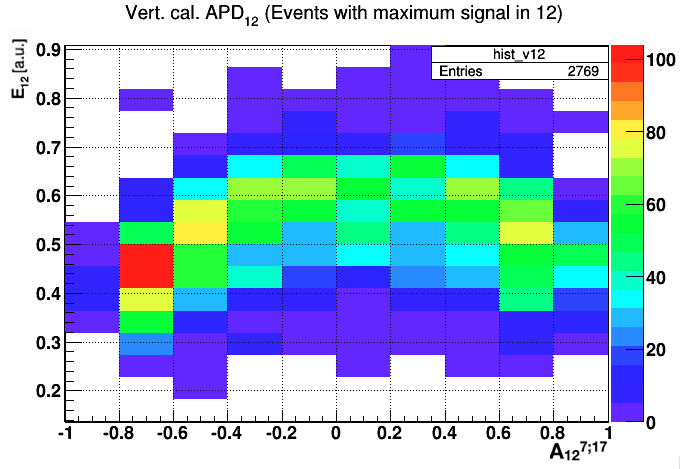}}
\hspace{0.001\linewidth}
\subfloat[]{\label{iso} \includegraphics[width=0.48\textwidth]{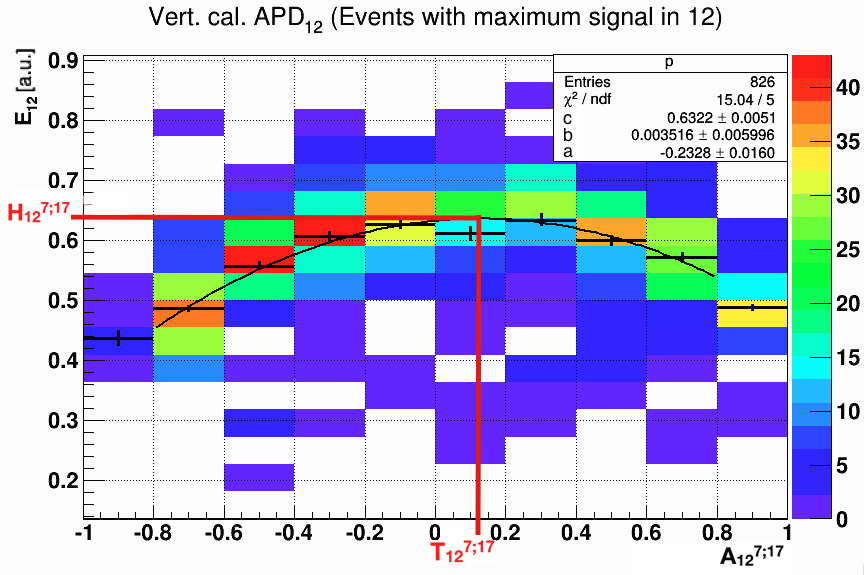}}
\caption{(a) $A_{12}^{7;17}$ (the {\it vertical asymmetry}) against the integrated signal $E_{12}$ when the maximum signal is at APD$_{12}$, (b) after the cut in the asymmetry in the perpendicular direction. Black error bars are the profile of the 2D histogram. The solid black line is the polynomial fit over the profile in the asymmetry range of $\pm0.8$. Solid red lines point to the two observables important for the inter calibration: $H_{12}^{7;17}$ and $T_{12}^{7;17}$.}
\label{fig}
\end{figure}

The behaviour of $A_i^{j;k}$ versus $E_i$ allows to characterize the charge sharing between APDs (see Fig.~\subref*{asy_cut}). A significant improvement of the $A_i^{j;k}$ versus $E_i$ distribution can be observed when a cut of $|A_{i}^{m,l}|<0.2$, where APD$_m$ and APD$_l$ are "perpendicular" to APD$_j$ and APD$_k$ 
ensuring that only events on the direct connection line between APD$_j$ and APD$_k$ are considered. Fig.~\subref*{iso} shows the improvement with this cut. 

If the gain of APD$_j$ and APD$_k$ is equal, $E_i$ should be maximum when the light sharing between APD$_j$ and APD$_k$ is the same ($E_j=E_k \Rightarrow A_i^{j;k}= 0$). Therefore, if the maximum value of $E_i$ is at $A_i^{j;k}\neq 0$ there is a gain difference between the two APDs. This deviation of  $A_i^{j;k}$ from zero, defined as $T_i^{j;k}$, is the first factor that provides quantitative information of this gain difference.

The second factor is called as $H_{i}^{j;k}$. It is the value of $E_i$ when $A_i^{j;k}=T_i^{j;k}$. This quantity can be derived from the $A_i^{j;k}$ versus the $E_i$ distribution as well. Assuming gain uniformity, it should be equal for all APDs. Therefore, the differences in $H_{i}^{j;k}$ provide information about the APD gains.

Fitting a second order polynomial: $E_i(A_i^{j,k})=a (A_i^{j,k})^2+b A_i^{j,k}+c$ to the average on each $x$-bin axis allows deriving the $T_i^{j;k}=-\frac{b}{2a}$ and $H_{i}^{j;k}=-\frac{b^2}{4a}+d$ (see Fig.\subref*{iso}). These factors can be determined using any pair of neighboring APDs if the symmetry between the charge cloud and the APDs position holds (see Fig.~\ref{apd_hor}). To obtain gain uniformity in the APDs, the relation between intercalibration factors, $c_i$, $T_i^{j;k}$ and $H_i^{j;k}$ is given by:

\begin{equation}\label{cal_A_E}
\frac{c_j}{c_k}=\frac{1+T_i^{j;k}}{1-T_i^{j;k}}\equiv \alpha_i^{j;k}~~~;~~~c_iH_{i}^{j;k}=\sum_{<j;k>} H_i^{j,k} \equiv \bar{H}_i^{j,k} 
\end{equation}

where $\sum_{<j;k>}$ represents the sum over all possible couplings. Since the system is overdetermined, we obtain the factors by minimizing eq.~\ref{eq_final_cal}. 

\begin{equation}\label{eq_final_cal}
\chi^2= \sum_{i=0}^{\#apds} \sum_{<j;k>} \left[ \left( \frac{\frac{c_{j}}{c_{k}}-\alpha_{i}^{j,k}}{\sigma_{\alpha_i^{j;k}}} \right)^2 + \beta \left(\frac{ \bar{H}_i^{j,k}-c_i H_i^{j,k} }{\sigma_{H_i^{j,k}}}\right)^2 \right]
\end{equation}

Here, $\sigma_{\alpha_i^{j;k}}$ and $\sigma_{H_i^{j,k}}$ are the errors of the $\alpha_i^{j;k}$ and $H_i^{j,k}$ observables respectively. The $\beta$ factor normalizes the value of $\sigma_{\alpha_i^{j;k}}$ and $\sigma_{H_i^{j,k}}$ in order to give similar weight to the two variables in the minimization. The minimization process can be run iteratively, thereby accurately determining the values the final calibration factors. The intercalibration was performed for each data set obtained under identical conditions (called run in the following). In sec.~\ref{impact} a study of the remaining miscalibration of the system is shown. It also explains how it can be partially corrected assuming illumination symmetry. 

\subsection{Energy and position reconstruction}
\label{Energyrecon}
The integrated signals $E_i$ of all APDs allow reconstructing the energy $Q$ released within the chamber and the position $\mathbf r=(x,y)$ of the event.
The simplest position estimation method is the centroid method. However, this method has the disadvantage that positions are biased towards the center of the APDs. For this reason we use in this study a method based on the comparison of measured and expected signals, $E_i$ and $QF_i(x,y)$ using a likelihood fit. $F_i(x,y)$ is the expected fraction of light in APD$_i$ (hereinafter called light-profile). This method allows estimating the position and energy of an event minimizing the following expression:

\begin{equation}\label{eq_chi2}
\chi^2=\sum_{i=0}^{\#apds}\sum_{j=0}^{\#apds}\left(E_i-QF_i(x,y)\right)\Sigma_{ij}^{-1}\left(E_j-QF_j(x,y)\right)
\end{equation}

Here, $\Sigma_{ij}$ is the covariance matrix which takes into account the uncertainty in the measured and expected signal in APD$_i$ and their correlation. The main uncertainty is due to the fluctuations of the measured signal which can be defined as $\Sigma_{ij}=<(E_i-<E_i>)><(E_j-<E_j>)>$. Then, the diagonal part $\Sigma_{ii}=\sigma_{E;i}^2$ takes into account the statistical fluctuations of primary signal, photoelectron detection, APD gain and the electronic noise in the device. Theoretically, it can be expressed as:

\begin{equation}
\Sigma_{ii}=\sigma_{E;i}^2=E_i^2\left[ \frac{W_{Xe}}{E_{xray}} \left( F_{Xe}+\frac{ENF_{i}-1+F_{Si}+\delta}{\eta \epsilon_{QE;i}\epsilon_{\Omega;i}} \right)\right]+\sigma_{ped;i}^2
\end{equation}

where $E_{xray}$ is the released energy inside the detector by the X-ray, $W_{Xe}$ is the average energy necessary to produce a electron-ion pair in Xe, $F_{Xe}$ is the Fano factor of xenon, $F_{Si}$ is the Fano factor of silicon, $\delta$ is the mean number of photoelectrons produced by each VUV incident photon, $\eta$ is the number of produced VUV photons per electron in the EL gap, $ENF_i$ is the Excess Noise Factor, $\epsilon_{QE;i}$ is the quantum efficiency, $\epsilon_{\Omega;i}$ is the geometrical efficiency and $\sigma_{ped;i}$ is the pedestal noise of APD$_i$. The off-diagonal term can be expressed as:

\begin{equation}
\Sigma_{ij}=E_iE_j\left[ \frac{W_{Xe}}{E_{xray}}F\right]+\sigma_{ped;i\times j} 
\end{equation}

where $\sigma_{ped;i\times j}$ is the covariance matrix of the pedestals. The computation of $\Sigma_{ij}$ requires knowledge of several quantities which depend on the conditions of the gas and the APDs during data taking. Furthermore, the factor $\epsilon_{\Omega;i}$ is very sensitive to the position of the X-ray interaction. The determination of this factor is done event by event. One way to determine the goodness of $\Sigma_{ij}$ is checking that the residuals of the minimization from eq.\ref{eq_chi2} follow a normal distribution with mean equal to zero and sigma equal to one. In standard conditions a mean of $0.02\pm 0.02$ and a sigma $1.09\pm 0.02$ are obtained for the 59.5 keV peak. This allows to estimate the theoretical values which serve as input to the Monte Carlo simulation.

This method relies on how well the light-profile model $F_i(x,y)$ describes the true light-profile. Several approaches to estimate the light-profile appear in the literature \cite{Neves200748}, \cite{Solovov201212}. In this study the problem was  tackled using two independent formalisms: an analytical and an iterative method, the two methods being completely independent of each other.\\

\subsection{Light-profile methods}
\label{sec_lightprof}
\begin{figure}[h!!]
\centering
\subfloat[]{\label{perfil_EL} \includegraphics[width=0.475\textwidth]{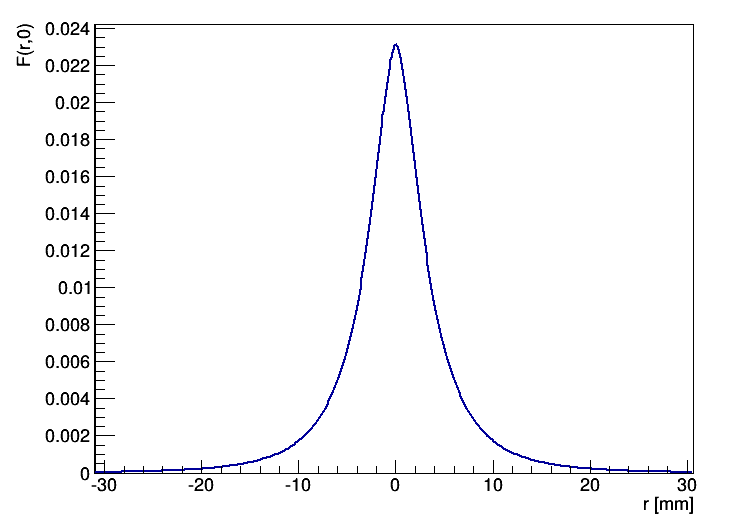}}
\hspace{0.001\linewidth}
\subfloat[]{\label{el_Sig} \includegraphics[width=0.485\textwidth]{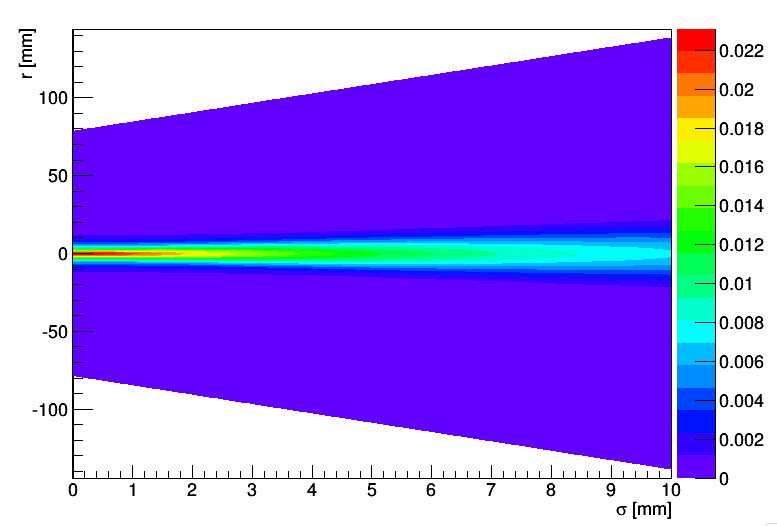}}
\caption{ (a) Simulated distribution of EL photons per electron in the readout plane ($F(r,0)$). Here, $r$ represents the distance between the electron position projected in the readout plane and the impact position of the EL photon in the readout plane and $F(r,0)$ is the fractions of photons in that position; (b) Expected light-profile for different $\sigma$ ($F(r,\sigma)$), where $\sigma$ represents the spread of the electron cloud in millimeters (in xenon gas under the conditions of our chamber, $\sigma=1.550\sqrt{z(cm)}~[mm]$, where $z$ is the drift distance), $r$ represents the distance from the center of the cloud axis in millimeters and the color represent the probability to have a EL photon in that position.}
\end{figure}

In the analytical method the values of $F_i(x,y)$ are calculated using macroscopic theoretical assumptions. First it is assumed that the charge deposition is point-like. This is not completely correct since electrons of 59.5 keV ejected from the gas atoms in a photoelectric absorption, have a range of about 5 mm in Xe at 3.8 bar \cite{NIST:Estar}. However, considering the large diffusion in Xe this assumption seems to be a good approximation as we will show in sec.~\ref{sec_results}. The diffusion of the charge cloud is also taken into account based on the diffusion coefficients obtained with the gas simulation program Magboltz \cite{Biagi1999234}. Furthermore, it is assumed that between the meshes the EL photons are emitted isotropically. The photon distribution in the APD plane mainly depends on the solid angle that an APD in the readout plane covers with respect to the center of the charge cloud  and on the transparency of the mesh.  An analytical model to describe the transparency of the grid, taking into account the radius and the pitch of the wires, was used. With these ingredients a simulation of the photon distribution produced by an electron drifting in the EL region can be obtained (see Fig.~\subref*{perfil_EL}). A convolution of this photon distribution with a Gaussian is calculated, in order to take into account the distribution of the electron cloud in the drift region (Fig.~\subref*{el_Sig}). Therefore, we add a parameter to the light-profile $F_i(x,y,\sigma)$ which is sensitive to the spread of the cloud. Ideally this degree of freedom should give us information about the $z$ positions of the events.

\begin{figure}[h!!]
\centering
\subfloat[]{\label{x0} \includegraphics[width=0.3\textwidth]{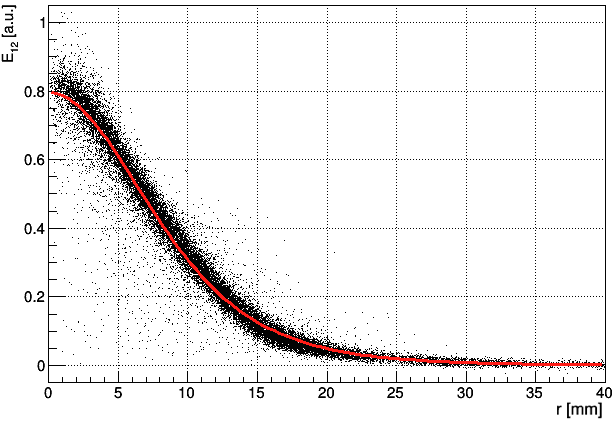}}
\hspace{0.001\linewidth}
\subfloat[]{\label{x1} \includegraphics[width=0.3\textwidth]{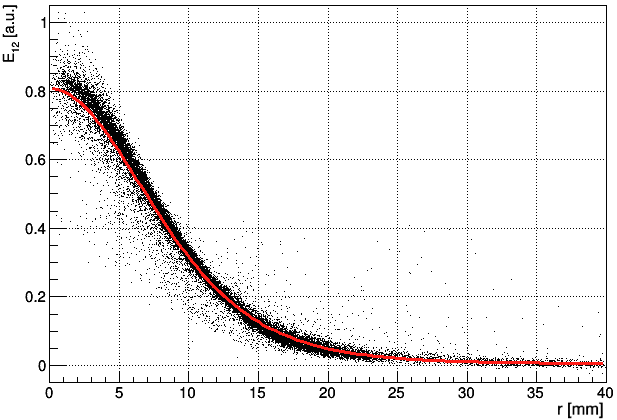}}
\hspace{0.001\linewidth}
\subfloat[]{\label{x4} \includegraphics[width=0.3\textwidth]{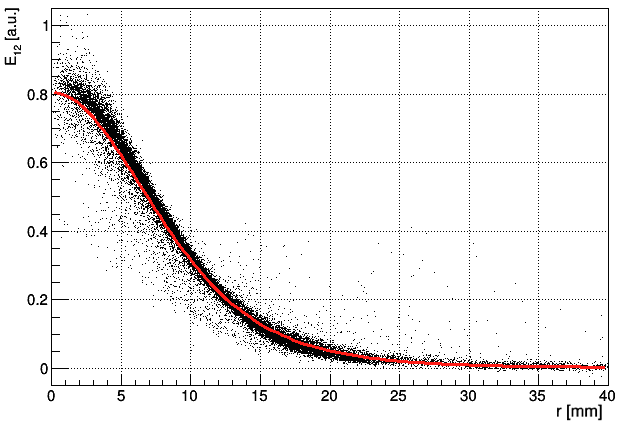}}
\caption{Signal of the central APD versus reconstructed distance (a) with the centroid method; (b) after one iteration; (c) after four iterations. The red line is the fit using eq.~3.6. 
Comparing (a) with (b) differences between the reconstructed events are clearly visible, especially in the central region. Comparing (b) with (c) the reconstructed events are very similar, which shows that convergence has been reached. The parameters from the fit also converge after several iterations.}
\end{figure}

The iterative method relies on the real response of the APD and was developed for a dual-phase xenon scintillation detector \cite{Solovov201212}. The $xy$ position is calculated on an event-by-event basis using the centroid method and the signal of each APD as a function of the $xy$ position was represented in a 3D histogram. These signals are then averaged bin-by-bin. A first approximation of the light-profile, $F_i(x,y)$, for each APD is obtained by fitting the averaged 3D histogram to a function defined as (see Fig.~\subref*{x0}):

\begin{equation} \label{eq_lightprofile}
F_i(x,y)=N_i \left(1+\frac{x^2}{\sigma_{i;x}\gamma_i}+\frac{y^2}{\sigma_{i;y}\gamma_i} \right)^{-\gamma_i}+B_i
\end{equation}

Here, $N_i$ is a normalization factor, $B_i$ is a term which takes into account the background for events far away from the APD, $\sigma_{i;x,y}$ reflects how peaked is the light-profile and $\gamma$ is the Cauchy-Gaussian factor. This fit can be used to do a better estimation of the position of the events using eq.~\ref{eq_chi2}. With this new position we can repeat the steps from the beginning to obtain a more reliable light-profile (see Fig.\subref*{x1}). This process can be repeated until convergence of the light-profile is found (see Fig.~\subref*{x4}).

\begin{figure}[h!!]
\centering
\includegraphics[width=0.5\textwidth]{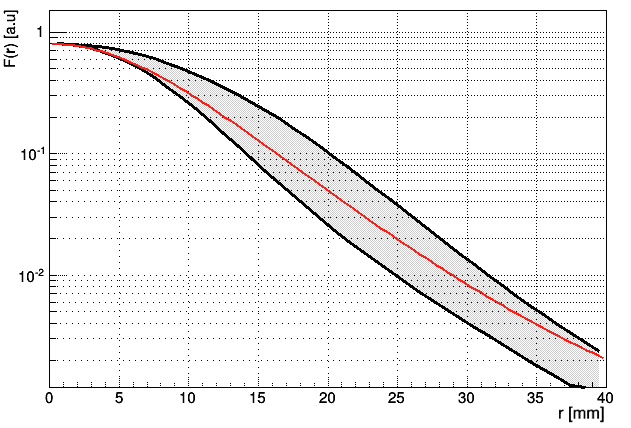}
\caption{Comparison between the results of the two light-profile methods. The red line is from the iterative method and the shadowed black area is from the analytical method limiting the parameter $5.5~mm<\sigma<7.5~mm$. }
\label{compare_log}
\end{figure}

This method also has drawbacks: it is very time consuming, the convergence strongly depends on the parametric function that is chosen (in our case the average number of iterations to converge is 4) and the position reconstruction for events centered on the APDs is more sensitive to fluctuations in the signals.
A comparison between the shapes of the light-profile obtained with the two methods can be seen in Fig.~\ref{compare_log}. It is obvious that the two give very similar results but they have two main differences: different values observed at the edges and the extra degree of freedom $\sigma$ introduced in the analytical method.

\subsection{Impact of Intercalibration and Light-profile determination} \label{impact}
\begin{figure}[h!!]
\centering
\includegraphics[width=.6\textwidth]{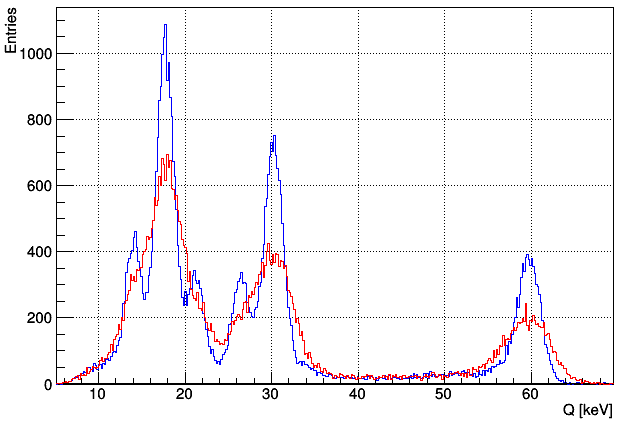}
\caption{$^{241}$Am spectrum at standard conditions (see Tab.~1). Blue and red lines are before ($R_E=9.5\%$) and after ($R_E=5.5\%$) event reconstruction applying  the analytical light-profile method and intercalibration. }
\label{spc_both}
\end{figure}

The importance of applying the APD intercalibration and of using the light-profile becomes obvious when one compares the spectrum before and after applying these corrections. Fig.~\ref{spc_both} shows the comparison. The energy resolution before intercalibration and light profile application is 9.5\% at 59.5 keV, improves to 8\% when only the light-profile is applied and to 5.5\% when intercalibration is applied as well. 

\begin{figure}[h!!]
\centering
\subfloat[]{\label{res_int} \includegraphics[width=0.5\textwidth]{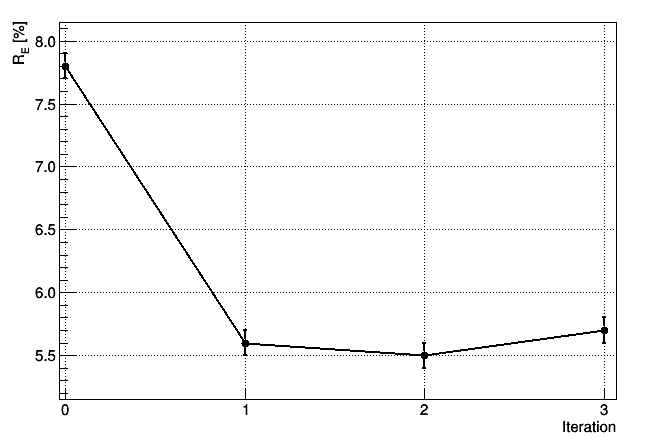}}
\caption{ (a) Resolution for 59.5 keV peak at standard conditions (see Tab.~1) for different numbers of iterations of the intercalibration ($0$ means no intercalibration). }
\end{figure}

Fig.~\subref*{res_int} shows the contribution of the intercalibration to this improvement. As one can see intercalibration improves energy resolution, from about 8\% (with light-profile event reconstruction) without any intercalibration to about 5.5\% after one iteration of the intercalibration procedure Ð at which point it is stable.  
 
\begin{figure}[h!!]
\centering
\subfloat[]{\label{QvsPhi} \includegraphics[width=0.475\textwidth]{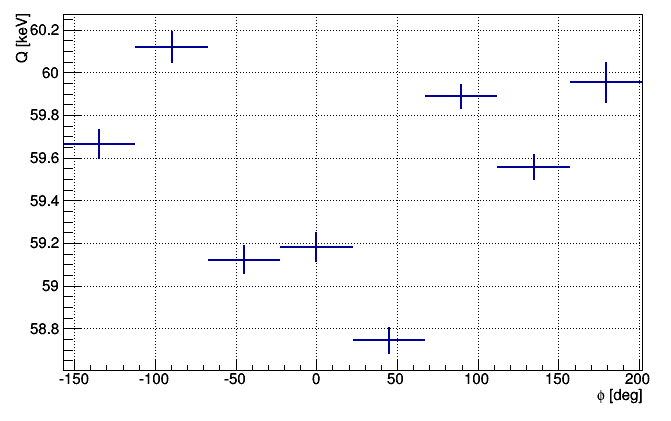}}
\hspace{0.001\linewidth}
\subfloat[]{\label{ResvsPhi} \includegraphics[width=0.485\textwidth]{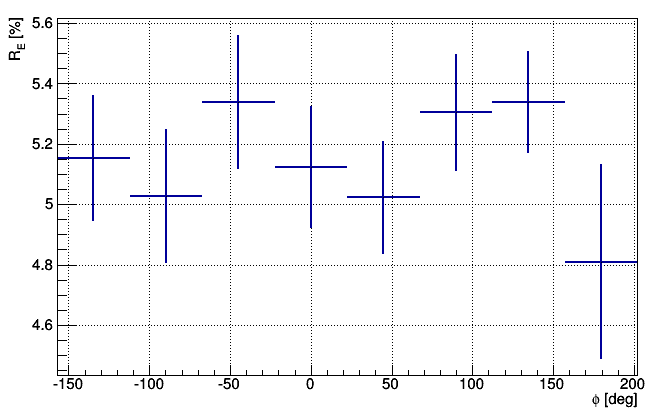}}
\caption{ (a) Position of the 59.5 keV peak against $\phi$ (defined as $\tan^{-1}\left( y/x \right)$, where $x$ and $y$ are the position coordinates in the readout plane) ; (b) Energy resolution of the 59.5 keV peak against $\phi$.}
\end{figure}

Comparing the results obtained with real data and Monte Carlo (MC) a discrepancy was found in the energy resolution of the 59.5 keV peak. The value obtained with real data is 5.5\% (see Fig.~\ref{spc_both}) while in the MC it is 5.2\% (see Fig.~\ref{Fig:QMC}). Furthermore, we observed that the 59.5 keV peak has a dependence in the position (see Fig.~\subref*{QvsPhi}) while the energy resolution does not (see Fig.~\subref*{ResvsPhi}).

This dependence can be explained with a remaining miscalibration of the APDs. Therefore, a tuning of the calibration factors is required to reduce this effect. In order to do that, calibration factors are multiplied by a factor that takes into account the shift of the 59.5 keV peak in the position of each APD. After this modification the RMS of the 59.5 keV peak as function of $\phi$ changes from 0.441 to 0.331 and the energy resolution becomes $5.3\pm0.1\%$, . compatible with MC value. Furthermore, there is a remaining miscalibration because this correction affects only partially the pattern observed in Fig.~\subref*{QvsPhi}. Based on the MC we expect that the remaining miscalibration is of the order of $10\%$.

\begin{figure}[h!!]
\centering
\subfloat[]{\label{s_it} \includegraphics[width=0.48\textwidth]{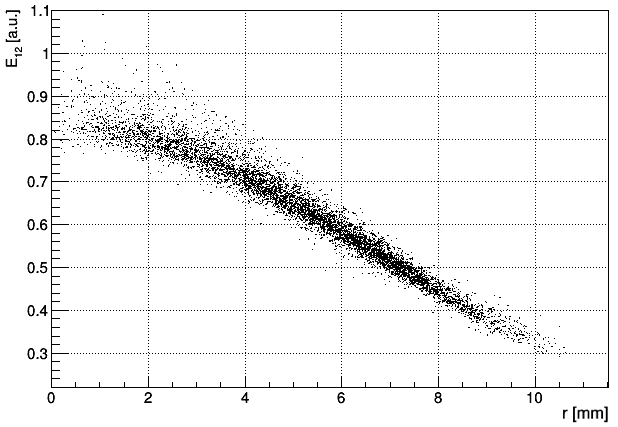}}
\hspace{0.001\linewidth}
\subfloat[]{\label{s_MC} \includegraphics[width=0.48\textwidth]{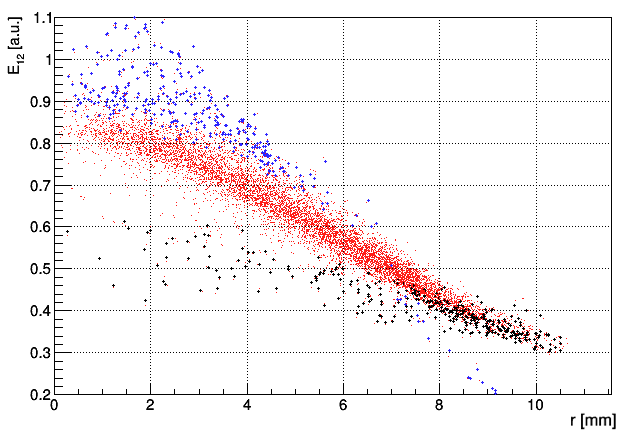}}
\caption{Signal versus reconstructed distance for 59.5 keV events: (a) iterative method; (b) analytical method (blue points are events with $\sigma<5.5~mm$ and black points $\sigma>7.5~mm$). }
\end{figure}

Fig.~\ref{res_old} compares the energy resolutions achieved with the two light-profile methods as a function of the APD bias voltage for 59.5 keV. It seems that the analytical method gives a better energy resolution especially with increasing APD gain and EL field. The differences are smaller in the low gain and low EL field cases. Despite these differences, the spectra obtained using the two methods are similar. We compare the signal of the central APD versus the reconstructed radial distance of the event from the center of the readout plane for the two methods. As shown in Fig.~\subref*{s_it}~and~\subref*{s_MC} a clear difference between the two methods is observed. The main difference arises from the $\sigma$ degree of freedom in the analytic case. Fig.~\subref*{s_it} clearly shows the main drawback of the iterative method, which arises from the difficulty of reconstructing the position of events with large signals in the central APD, because in this case there are large fluctuations.

\begin{figure}[h!!]
\centering
\includegraphics[width=0.8\textwidth]{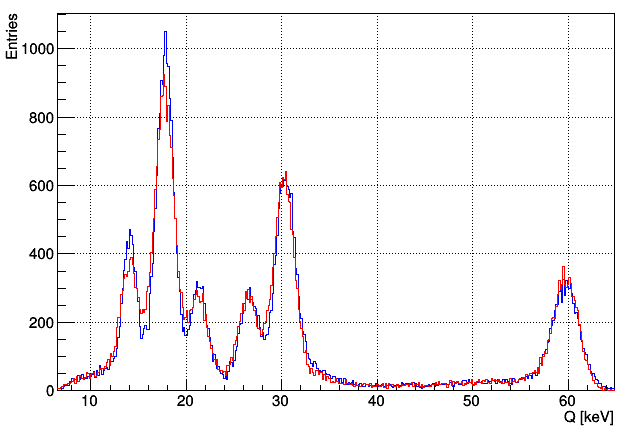}
\caption{ $^{241}$Am spectrum at standard conditions . Red and blue lines are analytical and iterative method respectively (analytical $R_E=5.3\pm0.1\%$ and iterative $R_E=6.0\pm0.1\%$).}
\label{nocut}
\end{figure}

It is interesting to analyze the energy spectrum using the two methods (see Fig.~\ref{nocut}). As stated in the previous paragraph, the two spectra have similar shape. The analytical light-profile method performs better in the 59.5 keV region, whereas a longer high energy tail appears with the iterative method. However in the low-energy region the performance is better using the iterative light-profile method.

\section{Characterization of the multi-APD EL-TPC}\label{sec_results}
The response of APD readout to EL signals  was characterized studying EL production, APD gain and energy resolution at 59.5 keV. We chose to operate at a drift field of 100 V/cm/bar, a value at which no dependence of EL and APD gain and energy resolution on the electric field strength was found. The temperature was monitored during data taking and kept stable within 1$^\circ$ C. All results given in this paper are for events with maximum signal in the central APD.

\subsection{EL and APD gain Dependencies}
\begin{figure}[h!!]
\centering
\includegraphics[width=0.6\textwidth]{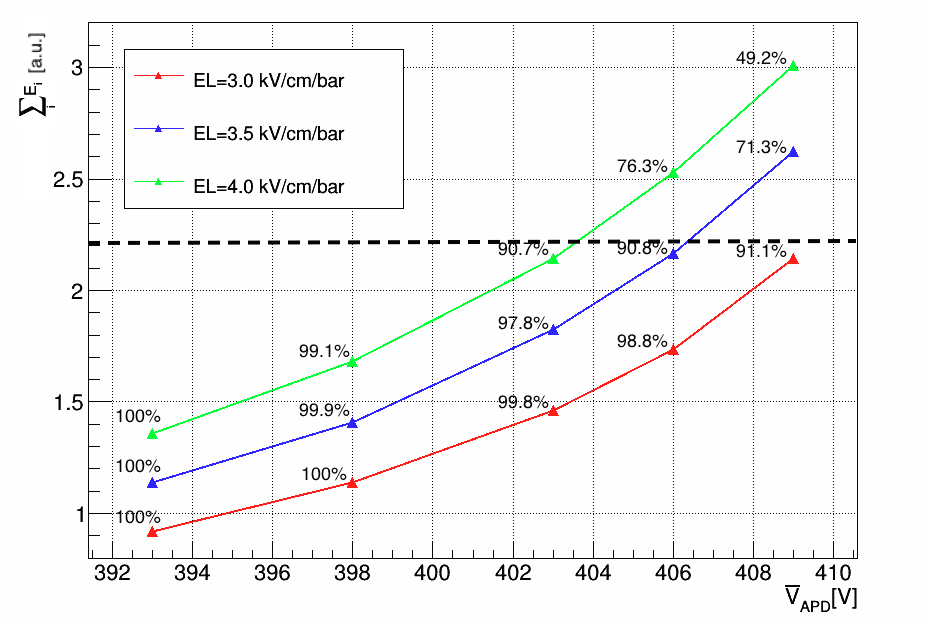}
\caption{ $\sum_i E_i$ for the 59.5 keV peak at $P=3.8~bar$ and $E_{drift}=100 Vcm^{-1}bar^{-1}$. The values next to the points are the percentage of 59.5 keV events that not saturate. The black dashed line defines the saturation cut.}
\label{gain}
\end{figure}

The sum of the integrated signals of all the APDs, $\sum_i E_i$, depends on the number of photons detected by all the APD$_i$ and their gain, which in turn depend on the voltage applied to the APDs and on the EL field. In Fig.~\ref{gain} the value of the mean of the 59.5 keV peak is plotted for three different EL field settings. The APD voltage given here, $\bar{V}_{APD}$, is the average voltage over all APDs.

The dynamic range of the ADCs is limited to twice the average APD voltage signal. Saturated signals are eliminated  by a cut on the events where the sum over all integrated signals, $\sum_i E_i$,  is greater than 2.2 (in the appropriate units - see Fig.~\ref{gain}).

\subsection{Energy resolution}

\begin{figure}[h!!]
\centering
\includegraphics[width=.6\textwidth]{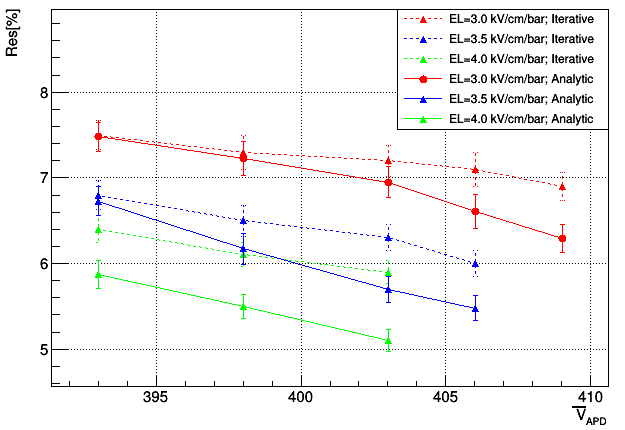}
\caption{Energy resolution of the 59.5 keV signal as function of the average APD voltage for 3 different EL fields and the two reconstruction methods.}
\label{res_old}
\end{figure}

In Figure ~\ref{res_old} the dependence of the energy resolution on $\bar{V}_{APD}$ is shown for three different EL fields.

A significant improvement of the energy resolution is obtained for reduced electric fields above $3~\textrm{kVcm}^{-1}\textrm{bar}^{-1}$ in the EL region. This is because at these EL field values the APD signal-to-noise ratio is enhanced.

\subsection{Linearity}

The linearity of the APDs as a function of the energy deposited in the detector was measured using the full $^{241}$Am X-ray spectrum, and reconstructing the energies with the iterative and analytic methods (see Figure~\ref{nocut}). First, a Gaussian is fitted to every peak seen in the spectrum. It must be noted, however, that the lowest-energy peak is very near to the pedestal cut, which makes less reliable the values obtained for it. Furthermore, all the peaks (except the one at 59.5 keV) are composed by a primary X-ray from $^{241}$Am and escape peaks (see sec.~\ref{sec_setup}). The expected values are calculated using a toy MC to simulate the experimental average of X-ray emission lines. Each peak is generated as a Gaussian with strength taken from its a-priori known value and a width from the experimental resolution. The result is then adjusted to a Gaussian and the central value used as the expected value.

\begin{figure}[h!!]
\centering
\subfloat[]{\label{linearity} \includegraphics[width=0.49\textwidth]{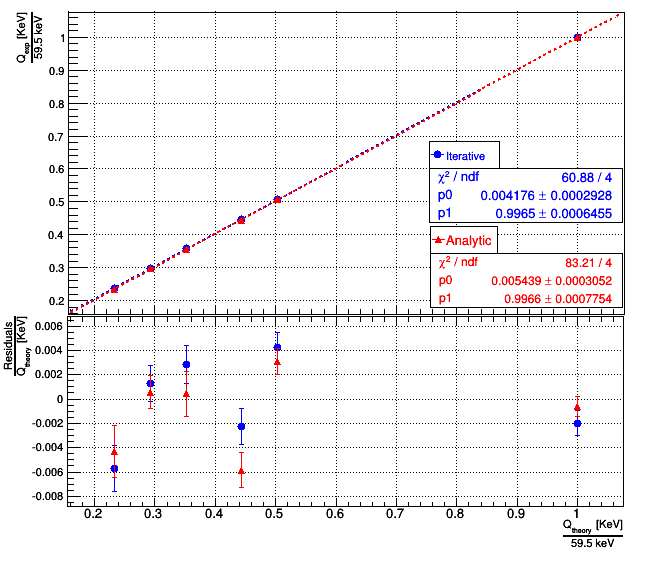}}
\hspace{0.001\linewidth}
\subfloat[]{\label{res_vs_e} \includegraphics[width=0.49\textwidth]{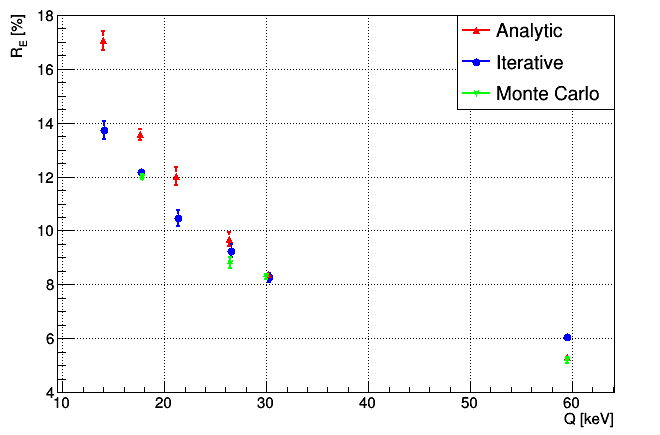}}
\caption{ (a) Linearity of the response to X-rays from the ${^{241}}$Am source referenced to the 59.5 keV signal. The linear fit residual are divided by the theoretical ratio; (b) Energy dependence of resolution.}
\end{figure}

On one hand, using the values of the X-ray peaks obtained from the fit we can compare the experimental values with the expected ones to study the linearity of the setup (Figure~\subref*{linearity}). A straight line is fitted to the ratio of the experimental peak values, normalized to the 59.5 keV peak, versus the expected peak positions. With both light-profile methods, the deviation of the slope from unity ($ 0.9965 \pm 0.0008$ ) and the offset from zero are small ($0.0054\pm0.0003$). This non-linearity is likely to be due the small signals in signal size in the APDs surrounding the one with the largest light deposition.  When these signals are near APD pedestals the reconstruction of the position in the $xy$ plane is affected. In turn, errors in reconstructing the position bias the energy reconstruction by introducing errors in evaluating the APD illumination. 

The availability of several X-rays of different energies allowed us to study the energy dependence of the resolution and to compare the results with the MC predictions (Fig.~\subref*{res_vs_e}). The MC describes well the shape of the distribution. The Monte Carlo simulations are described in the following section.  

\subsection{Position Resolution }

\begin{figure}[h!!]
\centering
\includegraphics[width=.47\textwidth]{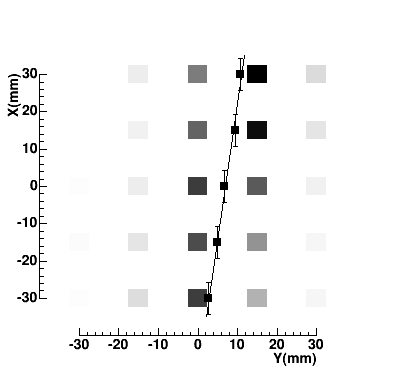}
\includegraphics[width=.47\textwidth]{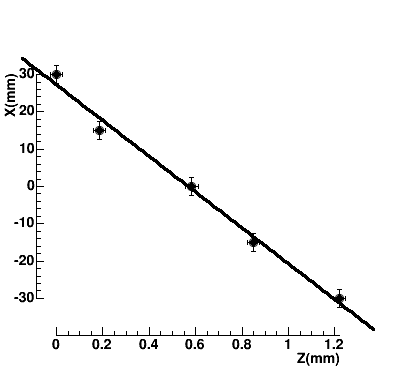}
\caption{ Cosmic muon event candidate in the APD plane (left) and
  drift direction (right). In the APD plane are shown the energy deposited 
  in each APD, in gray-scale, and the position in each APD row as obtained from
  the  likelihood fit and the linear fit. The error along the $x$ coordinate is
  the APD pitch divided by $\sqrt{12}$ and the error along the
  $y$ coordinate is the output of the likelihood fit.   The event display
  in the drift direction shows the position along the drift direction
  from the maximum of the waveform assuming the value of drift velocity
  to be 1.7~$mm/\mu s$. The error along the $x$ coordinate is
 the APD pitch divided by $\sqrt{12}$ and the error in $z$ 
 is fixed to be one unit of the ADC sampling clock. The $z$ coordinate of the track - in the
 drift direction - is zero-suppressed.}
\label{cosmics:Evt}
\end{figure}

The position resolution of the detector was measured using a sample of cosmic muons, recorded by triggering on the coincidence of the two scintillator counters. Due to the geometry of the scintillators the recorded tracks have a small angle to the direction of drift.  Fig.~\ref{cosmics:Evt} shows one cosmic event reconstructed in our detector. 

\begin{figure}[h!!]
\centering
\includegraphics[width=.7\textwidth]{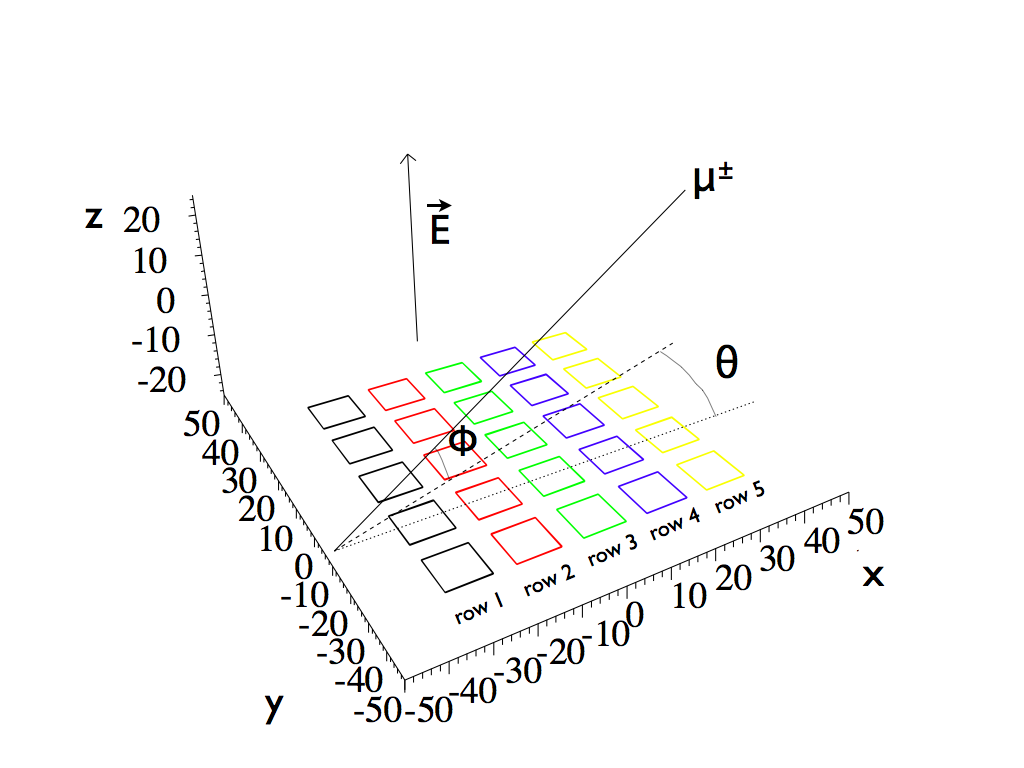}
\caption{ A track in the chamber. The track fit applies the likelihood method to
  individual rows. The coordinates are fitted to a straight line. The angle $\theta$ is that of the projection of the track in the tracking plane to the vertical axis.} 
\label{cosmics:EvtRep}
\end{figure}

    The APDs are grouped into rows of 5 in the horizontal plane
, see Fig.\ref{cosmics:EvtRep}. 
     The coordinates of the track are obtained independently for each of the five
     rows, using the same likelihood fit described for the $^{241}$~Am
    source but using the APDs in the same row. APD waveforms are integrated around the
    highest APD signal in a row and applying the same time window as that
    used for previous results.  The likelihood fit is performed row by
    row using the analytical light-profile, obtaining an energy
    deposition and transverse coordinate for each row.  The fit assumes
    that there is no fluctuation of ionization charge along the track. 
\footnote{ The track fit
      at higher angles can be improved by including the track angle
      in the prediction of the light sharing while performing the likelihood fit \cite{Carnegie:2004cu}. Here, we are only interested in the
      point resolution for vertical tracks, which is similar for both methods.  }

\begin{figure}[h!!]
\centering\subfloat[]{\label{cosmic:Angle} \includegraphics[width=0.5\textwidth]{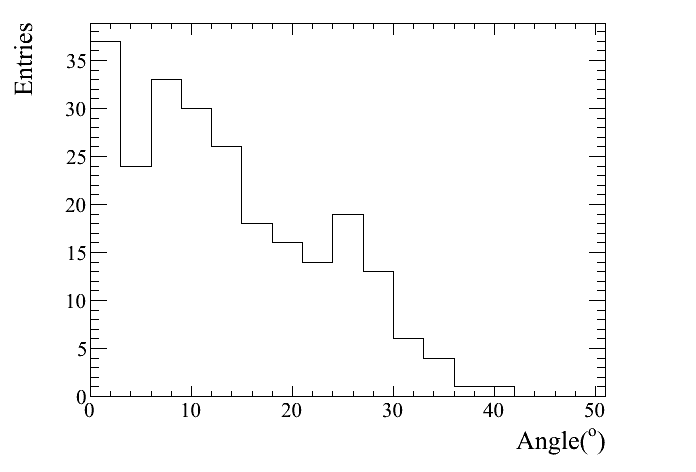}}
\subfloat[]{\label{cosmic:Charge} \includegraphics[width=0.5\textwidth]{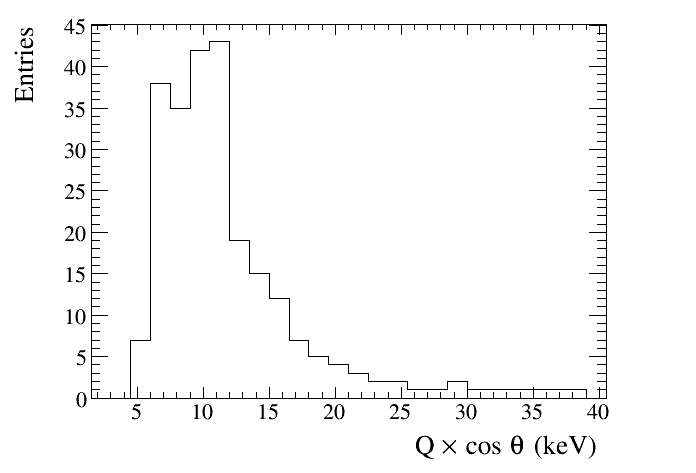}}
\caption{ a) Cosmic muon angular distribution with respect
  to the vertical axis. b) Charge deposited per row multiplied by the cosine of the angle of the 
  track to the vertical axis. }
\label{cosmic:ChargeAngle}
\end{figure}

    The track coordinates obtained for every row are then fitted to a straight line. The angular distribution of tracks is shown
    in Fig.~\ref{cosmic:Angle}. The reconstructed track charge 
   per row corrected by the 3D track angle ($\phi$ in
   Fig.~\ref{cosmics:EvtRep} ) is shown in
   Fig.~\ref{cosmic:Charge}. The deposited energy is calibrated using
   the 59.5~keV peak from the  $^{241}$~Am. The energy loss per unit track length at 3.8 bar, $dE/dx$, was calculated for cosmics using data from \cite{Biagi1999234}. Under these conditions and considering the APD dimensions $dE/dx$ is expected to be
 about 13.0 keV.  The mean value obtained from these data is ~12 keV, in good agreement with the predicted value. 

      The position resolution is computed as proposed in
      \cite{Carnegie:2004cu}.  For every APD row, we compute the track
      residual to the row coordinate for two cases: i) when that row's data are
      included in the linear fit, ii) when they are excluded from the
      linear fit.  The two distributions of residuals are 
      Gaussian-like. On one hand, the width of the residuals including
      the row in the fit ($\sigma^{all}_y$) is smaller than the real resolution due to the bias introduced by the row coordinate.
      On the other hand, the width of residuals when not using 
      the row in the linear fit ( $\sigma^{wo\,one}_y$) is larger because of the low number of points in the
      fit.  The best estimate of the point resolution\cite{Carnegie:2004cu} is obtained by taking
      the geometrical average of the width of the two residual
      distributions: 

     $$ \sigma_y = \sqrt{  \sigma^{all}_y  \sigma^{wo\,one}_y } $$

\begin{figure}[h!!]
\centering
\includegraphics[width=.5\textwidth]{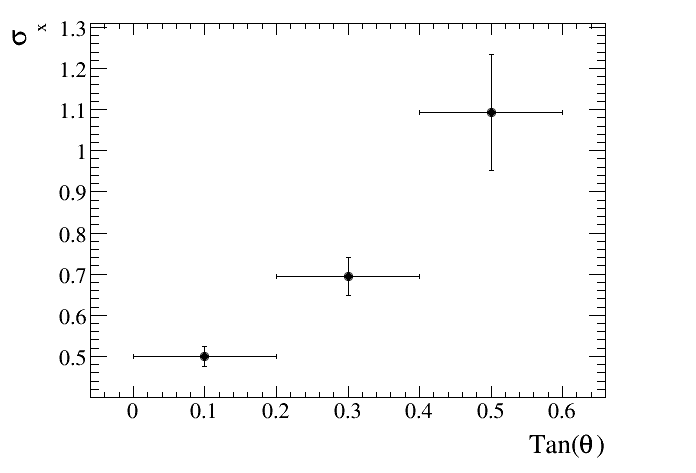}
\caption{ Position resolution as a function of the tangent of the track
  angle with respect to the vertical axis.  }
\label{cosmic:resvtan}
\end{figure}

        The position resolution as a function of the
        track angle projected into the tracking plane (see  Fig.~\ref{cosmics:EvtRep} ) 
        is shown in Fig.~\ref{cosmic:resvtan}.  The resolution is worse for
 tracks at a non-zero angle to the vertical due to the 
       light leaking from adjacent rows for tracks with different
       coordinates for each row.   We can estimate the optimal resolution for a point charge selecting the first angular
       bin, obtaining a value slightly lower than 0.5~mm. The spatial resolution is degraded  
       to 0.7~mm if one does not apply to the data the calibration described
       in sec.~\ref{APDcalibration}. 
       The sharing of light between several APDs with a 15~mm spacing improves the resolution from the result of the simple binary model, that would be $15/\sqrt{12} = 4.3$~mm, by a factor of $\approx$~8. Of course with more APDs per unit surface more APDs would receive light and altogether more light would be detected, with a further improvement of the resolution. 

\section{Simulation of signal generation in the multi-APD EL-TPC}
\label{sec_mc}

In order to better understand the energy and position resolutions measured in the laboratory and to be in a position to extrapolate them to different experimental conditions we developed a full Monte Carlo simulation of the X-ray interaction, electron transport, EL photon production and detection. The simulation steps are described in this section. \\ 
First, using the Geant4 code\cite{Agostinelli2003250}, the propagation and interaction of a 60 keV X-ray in a volume of xenon at 3.8 bars are simulated. This is done with the Penelope package \cite{Baro199531} that describes detailed features of the interactions of low energy gammas with matter.  Geant4 provides a set of point energy depositions that are treated independently. The first step computes the number of primary electrons using the formula: 

\begin{equation}
N_e= \frac{E_{xray}}{W_{Xe}}
\end{equation}

where $E_{xray}$ is the energy released in one step of charged particle transportation within the detector's active volume, $W_{Xe}$ is the average energy to produce an electron-ion pair in xenon, taken to be $21.6$~eV \cite{DoCarmo2008}. The number of electrons per step is smeared with a Gaussian distribution centered at $N_e$ and sigma equal to \cite{Borges1999321}.

\begin{equation}
\sigma_{N_e}=\sqrt{N_e F} 
\end{equation}

where $F$ is the Fano factor. Different Fano factors for xenon are given in the literature. We selected the value of $0.14$ \cite{DoCarmo2008} which is obtained in conditions close to our experimental setup. Next, the electrons are propagated to the electroluminescence mesh taking into account transverse diffusion. We assume 100\% transparency for the electrons entering the EL gap. The number of photons produced in the EL gap per electron is well described by the experimental parametrization \cite{Monteiro:2007vz}:

\begin{equation}
\eta=140 \left( \frac{\Delta V}{p\Delta z}-0.83 \right) p\Delta z
\end{equation}

Where $\Delta V$ is the voltage difference between the two meshes in the EL region, $p$ is the gas pressure within the chamber and $\Delta z$ is the distance between the EL meshes. This expression is valid for reduced electric fields under 8 kVcm$^{-1}$bar$^{-1}$. Under the detector's operating conditions the value of $\eta$ is approximately $1100$.  The transmission of the mesh for photons and the APD acceptance coverage have been simulated with the analytical model used for the light profile as described in sec.~\ref{sec_lightprof}. The average photon acceptance of the array of 25 APDs is computed to be between 0.9\%  and 1.0\% depending on the location of the 60 keV deposition within the chamber's active volume.

Each electroluminescence photon creates an average of 2 electrons in the silicon because the energy required to create an electron-hole pair in the silicon is  3.6~eV \cite{Beringer:1900zz} while the energy of the EL photon is 7.2~eV. This value is fluctuated assuming a Fano factor of 0.1 \cite{Beringer:1900zz}. The number of electrons per photon impinging on the APD was measured to be  $0.69\pm0.15$ \cite{Lux201211}. This value must be reduced by the average number of electrons per photon conversion.  The number of photons converted in the APD is fluctuated assuming a Poisson error and then multiplied by the number of electrons generated in the conversion. The dispersion introduced by the gain fluctuations in the APD, i.e. the Excess Noise Factor (ENF), is not well measured because it depends on the device and operating conditions. The typical value for APDs is around 2.0 depending on the assumptions \cite{Ikagawa2003671}.  We set the ENF to 2.0 for all the APDs. This is an approximation because APDs are operated at different bias voltages and therefore different values of the ENF can be expected. Electron transport, photon production and detection are computed for all ionization points predicted by Geant4. This allows taking into account possible geometrical effects due to the finite spatial extent of the energy deposition in the gas. 

Finally, to simulate remaining APD miscalibrations, for each APD the signal is multiplied by a random number chosen in the range $[1.-\delta_{cal},1.+\delta_{cal}]$. The $\delta_{cal}$ value was estimated to be about $0.10$ from the remaining variance of the maximum value, $H^{j;k}_i$, (see sec.~\ref{APDcalibration}) after applying the calibration.  Once the total energy per APD is recorded, we simulate the effect of the electronic noise taking into account the correlations between APDs as indicated from the data themselves, as shown in section \ref{SignalTreatment}. The intercalibration value and the value of the pedestal width from the experimental setup allows us to predict the typical pedestal width of $\approx 100$~photoelectrons.

\begin{figure}[h!!]
\centering
\includegraphics[width=0.6\textwidth]{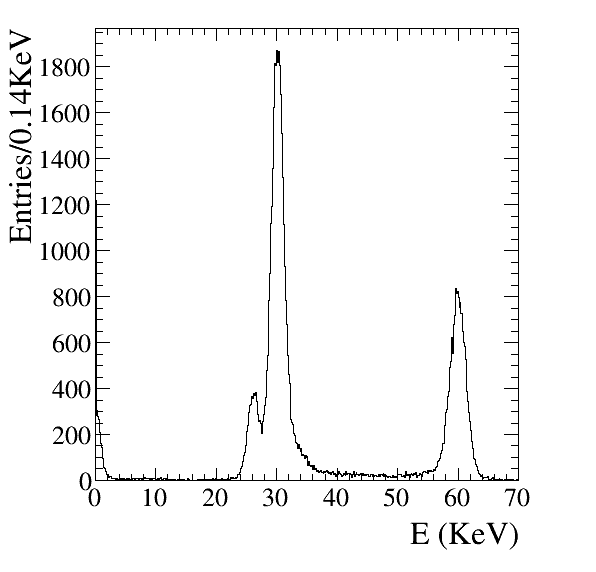}
\caption{ Spectrum obtained from the full simulation of a 60~keV photons in a 3.8~bar xenon TPC. The result corresponds to the calculations done with $F=0.14$, $\delta_{cal} = 0.10, ENF=2.0$ and including electronics noise. }
\label{Fig:QMC}
\end{figure}

The APD signals are fitted using the likelihood method described in section~\ref{Energyrecon}, providing an estimate of the energy resolution. The spectrum obtained with this model is shown in Fig.~\ref{Fig:QMC}. The figure shows the main features of the interaction of 60 keV photons with the xenon, with two escape peaks. For 60~keV gammas the model predicts a resolution of  $5.18\pm0.05$\% (FWHM) for the nominal parameter set, which follows the current state-of-the-art in the literature. Values of the resolution obtained by varying this parameter set are shown in Table \ref{Tab:MCRes}. The predicted resolution is close to the experimental value from the 59.5 keV gamma rays.  The main contribution to the energy resolution is due the electronic noise, as a consequence of the low photon acceptance ( $\approx$ 1\%).  We have also evaluated the contribution of the dispersion of the energy deposition within the detector by simulating that all the energy is released at one point. In this case the energy resolution is $5.0\%$ (FWHM), lower than the previous result, and showing that the spatial dispersion of the energy deposition contributes to the final resolution. The energy resolution of the escape peaks around 30 keV and 26~keV were found to be $8.29\pm0.06$\% and $8.84\pm0.17$\% respectively. In addition we have simulated with the same parameters settings a 17.8~keV gamma source, obtaining a 12\% (FWHM) energy resolution. All these numbers are compared with the experimental results in Fig.~\ref{res_vs_e}.

\begin{table}[h!!]
\begin{center}
\begin{footnotesize}
\begin{tabular}{| c|c|c|c|c|c|c|}
\hline
  Energy (keV)  & Nominal                   & $F=0.29$                        & $\delta_{cal} = 0.00$  & Noise off   & $ENF=3.0$ &  $ENF = 1.8 $ \\

\hline
60.0               & $5.2\%$                     & $5.5\%$                           & $4.9\%$                    & $4.5\%$    &     $5.6\%$               &  $5.1\%$ \\
\hline
\end{tabular}
\end{footnotesize}
\end{center}
\caption{Predicted energy resolution (FWHM) changing one parameter at a time and fixing the rest to the nominal values ($F=0.14$, $\delta_{cal} = 0.10, ENF=2.0$ and noise on)  as described in the text. }
\label{Tab:MCRes}
\end{table}

Taking advantage of the satisfactory agreement of the model with the experimental results, we have explored the potential resolution of several APDs configurations separated by 5.1~mm (almost full coverage)  and 10~mm - instead of the 15~mm in this setup - and increasing the electroluminescence photon yield. The predicted energy resolutions are tabulated in Tab.~\ref{Tab:Compact}. Although there might be subleading contributions to the energy resolution that are not included in this model and might account for the difference between the experimental results and the model, it is shown that the actual energy resolution can be improved significantly by increasing the electroluminescence photon statistics. This can be achieved with a larger photon acceptance (see last column in Table \ref{Tab:Compact} ) and by increasing the electroluminescence yield.  The best resolution obtained is excellent (2.3\%) and close to the expectation from fluctuations in the yield of primary electrons ($FWHM=1.7\%$ with a Fano factor $F$ = 0.14). The values for 5.1~mm pitch and large $\gamma/e^-$ show similar results due to the dominance of other effects like APD miscalibration, the dispersion of the energy deposition in the detector, Fano factors, etc. The achieved energy resolution shows the potential of this type of readout technology for combined tracking/positioning and calorimetry measurements.

\begin{table}[h!!]
\begin{center}
\begin{footnotesize}
\begin{tabular}{|l|c|c|c||c|}
\hline
                            &  1100 $\gamma/e^-$ &  2200 $\gamma/e^-$ &   4400 $\gamma/e^-$ & APD Illumination\\
\hline
 15 mm (nominal) & $5.2\%$                     &  $3.9\%$                       &  $3.2\%$                      &  $\approx 11\%$ \\
\hline
 10 mm                & $3.8\%$                      &  $3.0\%$                      &  $2.8\%$                     & $\approx 24\%$\\
\hline 
 5.1 mm               &  $2.7\%$                     &  $2.2\%$                      &  $2.3\%$                     & $\approx 67\%$ \\
\hline
\end{tabular}
\end{footnotesize}
\end{center}
\caption{Predicted energy resolution (FWHM) for different pitches of the APD array (rows) and electroluminescence photon gains (columns). The last column shows the fraction of photons falling within the APD acceptance normalized to the number of photons transmitted through the lower mesh for 25 APDs in the readout plane. The values for a pitch of 5.1 mm and 2200 and 4400 $\gamma/e^-$ are in agreement within the error of the result. }
\label{Tab:Compact}
\end{table}

Based on the satisfactory agreement of the MC prediction with the experimental result and in view of potential applications to a Compton camera we have explored the response of the detector to higher-energy gammas. We consider slightly different experimental conditions that would be more appropriate for this purpose:  10 bar xenon gas and a fully packed array of APDs with a pitch of 5.1 mm. We consider a 360~keV gamma source with the Compton edge at 213~keV and a minimal second photon energy of 150~keV. To evaluate this scenario we consider three typical energies: 210~keV, 140~keV, 50~keV and the 30~keV corresponding to the xenon $K_{\alpha}$ escape peak. The energy resolutions, to be compared with a typical solid state detector resolution of 2.5\% (FWHM) at 81 keV \cite{Takeda5076099},  are tabulated in Tab.\ref{Tab:ComptonE}. An angular resolution of  0.7$^{\circ}$ (FWHM) is estimated from the average flight path of a 140 keV photon at 10 bar (19.9~cm) and the error obtained for the position resolution from the cosmic analysis (0.5~mm). The angular resolution will be dominated by Doppler broadening which is estimated to be 2$^{\circ}$ (FWHM) \cite{Zoglauer2002}.

\begin{table}[h!!]
\begin{center}
\begin{footnotesize}
\begin{tabular}{|c|c|}
\hline
   Photon energy   &  Energy resolution (FWHM) \\ 
\hline
 210   keV         &  $1.3\%$          \\ 
\hline
 140   keV          & $1.5\%$           \\ 
\hline 
  50    keV          &  $2.6\%$           \\ 
\hline 
  30    keV          &  $3.2\%$           \\ 
\hline
\end{tabular}
\end{footnotesize}
\end{center}
\caption{Expected energy resolutions (FWHM) for photon energies relevant for a Compton camera with a 360~keV gamma source. }
\label{Tab:ComptonE}
\end{table}

\section{Conclusions}\label{sec_conclusions}
In this paper we presented a study of the performance of a multi-APD electroluminescence TPC filled with pure xenon at 3.8 bar which achieved 
an energy resolution as good as 5.3$\pm$0.1\% FWHM for the 
59.5 keV peak from $^{241}$Am. It was shown that the intercalibration of the APDs is crucial for the performance of the readout because it improves the energy resolution 
from 8\% to 5.5\%.  Correction for the variations remaining after  intercalibration yields the previously mentioned energy resolution of 5.3\%. As a consequence
for larger detectors an improved calibration method will have to be developed.  
In addition the point resolution of the instrument was measured using cosmic rays, obtaining point resolutions of about 0.5 mm for APDs with a pitch of 15 mm
between them. \\
A Monte Carlo simulation based on Geant4 and Penelope was developed to better understand the result on the energy resolution 
and to provide a tool to extrapolate to operation with different APDs, readout geometries and higher pressures. For the operating parameters
used for the measurements the simulation predicts an achievable energy resolution of 5.2\% FWHM for 60 keV, in good agreement with the experimental result. \\
Extrapolating to higher readout densities and higher pressures reveals that significantly better energy
resolutions, down to 3\% at 59.5 keV with a 10 mm pitch between APDs and 10-15 bar could be achieved. These results confirm that a multi-APD EL TPC is an interesting option in applications such as Compton and gamma cameras where excellent energy and spatial resolution are required. Increasing further the fraction of the detection plane covered with detectors might push the energy resolution down to 2.2-2.3\% for 59.5 keV. 
    
\acknowledgments
The authors acknowledge the support received from the Ministerio de Ciencia e Innovaci\'{o}n under grants FPA 2011-29823-C02-02, Consolider Ingenio Project CSD2008-0037 (CUP), Consolider Ingenio Project CSD-2007-00042 (CPAN) and Centro de Excelencia Severo Ochoa SEV-2012-0234, some of which include ERDF funds from the European Union. We would like to thank  M. Ieva for fruitful discussions during the development of this research and J. Gaweda for technical support. Finally, we would like to thank M.Cavalli-Sforza for his editing and English proof reading.



\bibliography{EL_DB}{}

\bibliographystyle{JHEP}

\end{document}